\input preprint.sty
\input epsf
\pptstyle
\letter{Trans--Planckian modes, back--reaction, and the Hawking process}

\author{ Adam D Helfer}

\address{Department of Mathematics, Mathematical Sciences Building, 
University of Missouri, Columbia, Missouri 65211, U.S.A.}

\jl{6}

\beginabstract
Hawking's prediction of black--hole evaporation depends on the
application of known physics to fantastically high energies --- well
beyond the Planck scale.  Here, I show that before these extreme
regimes are reached, another physical effect will intervene:  the
quantum backreaction on the collapsing matter and its effect on the
geometry through which the quantum fields propagate.  These effects are
estimated by a simple thought experiment.  When this is done,
it appears that there are no matrix elements allowing the emission of
Hawking quanta:  black holes do not radiate.
\endabstract

\pacs{04.70.Dy, 04.62.+v, 03.70.+k}

\date



\def\lesssim{{\mathrel{<\atop\sim}}}
\def\gtrsim{{\mathrel{>\atop\sim}}}
\def\CMP{{Commun. Math. Phys.}}
\def\scri{{\cal I}}
\newcount\EEK
\EEK=0
\def\eek{\global\advance\EEK by 1\eqno(\the\EEK )}

\nosections 

Hawking (1974, 1975) famously studied quantum fields in the presence of
a gravitationally collapsing object, and predicted that black holes are
not in fact black, but radiate with thermal spectra and eventually
explode.  This work raised many issues, and forced the development
of far deeper understandings of quantum field theory in curved
space--time than had previously been achieved.  Whether black holes
radiate or not, these deeper foundations seem secure.  It is no
exaggeration to say that almost every important paper in the past
twenty--five years on quantum field theory in curved space--time has
Hawking's papers as antecedents.

The strength of the argument for black--hole evaporation is at best
dubious.  The main difficulty, which was recognized almost from the
outset, is that Hawking's analysis relies on the application of known
physics to fantastically high energies, well beyond the Planck scale.
In spite of this, the prediction of black--hole evaporation has often
been regarded as a cornerstone of the theory, to such an extent that
attempts to quantize gravity have been judged by whether they can
reproduce Hawking's prediction in appropriate regimes (Rovelli 1996,
Strominger and Vafa 1996, Balbinot and Fabbri 1999, Kummer and
Vassilevich 1999; see however Belinski 1995 for another
view).  So it is worthwhile reconsidering Hawking's
argument, not just for itself, but for its implications for more
ambitious theories.

The kernel of the trans--Planckian problem is this.  The predicted
spectrum corresponds to a temperature $T_{\rm H}=\hbar c^3 /8\pi Gm$
(where $m$ is the mass), and so to wavelengths of characteristic size
$\sim 8\pi Gm/c^2$.  However, the field modes from which these arise
have been exponentially red--shifted as the modes propagate away from
the collapsed object, so that the original wave--lengths they
corresponded to, in the distant past, are $\sim (Gm/c^2)\exp -u/(4m)$, where
$u$ is the retarded time.  For a solar--mass black hole, the scale $4Gm/c^3$
is on the order of milliseconds, so in a fraction of a second the
energies of the modes have passed any known physical
scale --- including the mass of the universe!

No resolution of the trans--Planckian problem is known.  There have
been a number of ingenious investigations which have supposed that the
effect of trans--Planckian frequencies on the field theory might be modeled
by a dispersive propagation of the quantum field (Jacobson 1991,
1993, Brout et al. 1995, Unruh 1995).  However, all of these are
guesses at what physics at very high energies will be, all
significantly alter the propagation assumed by Hawking, and none seems
able to reproduce Hawking's result without introducing trans--Planckian
wave--numbers.  It has also sometimes been suggested that, because of
the very beautiful form of Hawking's results, there ought to be some
sort of general invariance arguments leading to them, independent of
trans--Planckian physics.  However, at present no such argument is
known.

In this letter, I will show that before the trans--Planckian regime is
reached, another physical effect will have to be considered:  the
quantum back--reaction on the collapsing matter, and its effect on the
space--time geometry through which the quantum field propagates.  That
this should need to be considered is at first surprising, since for
almost all purposes gravitational fields can be adequately computed by
considering their sources as classical objects.  But the Hawking
mechanism is not ordinary physics.  It is a tiny quantum effect, and it
turns out to be crucially influenced by quantum complementarity issues
involving the geometry and the collapsing matter.

In what follows, I shall consider only the case of spherical symmetry
explicitly, but it will be evident that the physical arguments are quite
general and should apply more broadly.  The conventions are those of
Penrose and Rindler (1984--6), and of Schweber (1961).  The metric has
signature ${}+{}-{}-{}-{}$.  For the most part, factors of $G$, $\hbar$
and $c$ are given explicitly, but these are omitted where the
expressions become too cumbersome.

In the space--time exterior to the collapsing body, we have the
Schwarzschild metric in the familar form
$$\d s^2=(1-2m/r)\d t^2-(1-2m/r)^{-1}\d r^2 -r^2\d\theta ^2
  -r^2\sin ^2\theta \d\varphi ^2\, ,\eek$$
and we also make use of the advanced and retarded time coordinates
$$v=t+r_*\qquad\hbox{and}\qquad u=t-r_*\, ,\eek$$
where $r_*=r-2m+2m\log ((r-2m)/2m)$.

The general scheme of Hawking's computation is this.  A massless quantum
field $\hat\phi$ is to be investigated on the space--time corresponding
to a gravitationally collapsing object.  The state of the field in the
far past is quiescent, say vacuum for simplicity.  In principle, then, one
should work out the field operators in the far future in terms of those
in the far past, and from this one can read off the particle--content,
stress--energy, etc., of the state in the far future.

The leading contribution to the field operators in the future far from
the object and after collapse has substantially occurred will be the
geometric--optics approximation.  We may write this as
$${\hat\phi}^0_{\rm f}(u)={\hat\phi}^0_{\rm p}\bigl( v(u)\bigr)\,
,\eek$$\xdef\fmap{\the\EEK}%
where the subscripts ${\rm f}$, ${\rm p}$ stand for future and past, the
superscript indicates that the fields have been conformally rescaled to
attain finite limits at $\scri ^\pm$, and $u\mapsto v(u)$ is the mapping
of surfaces of constant phase, from $u={\rm const}$ in the future to
$v=v(u)$ in the past.

Equation (\fmap ) is remarkable in that it has the same form as a ``moving
mirror'' model:  a massless field propagating in two--dimensional
Minkowski space and reflecting from a perfect mirror whose trajectory is
given by $v=v(u)$ in null coordinates (Fulling and Davies 1976, Davies
and Fulling 1977).
(The moving mirror models were
developed after Hawking's work, however, and in part to help understand
it.)  Using standard formulas from these models, we may read off the
particle--content and stress--energy.  We find
$$\langle {\widehat T}_{uu}^{\rm ren}\rangle =
  (12\pi r^2)^{-1}\hbar \left( -(1/2) (v''/v')' +(1/4)(v''/v')^2\right)
  \, .\eek$$\xdef\stres{\the\EEK}%
Note that the key quantity that enters is the fractional acceleration
$v''/v'$.  This is the center of the Hawking mechanism.  The
aspect of the space--time geometry which controls the Hawking mechanism
is $v''/v'$, and measurements of Hawking quanta are essentially probes
of this geometric quantity.

According to classical collapse theory, one has, at late retarded times
$$v(u)\sim -(Gm/c^3)\exp -u/(4m)\, ,\eek$$\xdef\exrel{\the\EEK}%
where the prefactor $Gm/c^3$ has been inserted for dimensional reasons only;
one has no control over the constant in front of the exponential at this
level of analysis.  From this relation and equation (\stres ), one
immediately has $\langle {\widehat T}_{uu}^{\rm ren}\rangle =
(48\pi r^2)^{-1}\hbar (c^3/4Gm)^2$.  This is one of Hawking's predictions,
and others can be similarly recovered.

The relation (\exrel ) is the one which gives rise to the trans--Planckian
problems.  But exponential relations like (\exrel ) are never accepted
unreservedly in physics.  While they may apply within a particular
model, no model holds at arbitrarily small scales, and eventually one
must ask what aspect of the model breaks down.  Here we shall see that
there are quantum limitations.

We shall show that it is necessary to consider the quantum
back--reaction on the collapsing matter and the space--time geometry.
Since this is precisely a question of how geometry is affected by
quantization, and we have at present no reliable theory of quantum
gravity, any investigation along these lines involves some
speculation.  However, here we are not concerned with
gravitational fields which are locally very strong in any invariant
sense, nor (it will turn out) with Planck--scale physics, so it seems
reasonable that we should be able to apply conventional physical
principles to understand them.

Accordingly, we shall consider a simple thought--experiment to
measure $v''/v'$.  The idea is direct.  We imagine sending massless
particles of a given frequency into the collapsing body, slightly before
the point where they would inevitably be captured.  The particles
emerge, and we measure their red--shifts, or more precisely certain
ratios of these, to compute $v''/v'$.  See figure 1.

In this experiment, and in any real experiment, one does not measure
$v''/v'$ at an instant; one measures an average of it over some finite
time.  For our present purposes, the time scale of interest is $\sim Gm/c^3$,
the characteristic time for the emission of a Hawking quantum.  If
$v''/v'$ is to be measured over such a time, then the quanta used must,
on arrival, have frequencies $\gtrsim c^3/(Gm)$.  However, this means that
the initial quanta must have had frequencies of order $\gtrsim
(c^3/Gm)(v')^{-1}\sim (c^3/Gm)\exp +u/(4m)$.  We thus very quickly pass
a point where
$$(\hbar c^3 /Gm)(v')^{-1}\gtrsim mc^2\, ,\eek$$\xdef\scd{\the\EEK}%
that is, where the energies of the incoming quanta must exceed that
of the collapsing object.  By this point, measurements of $v''/v'$ of
sufficient precision to correspond to the geometry determining the
Hawking phenomenon would require disturbances of the energy of the same
order as the energy of the collapsing object itself, and the whole
notion of probing a background geometry has clearly broken down.  

It is precisely the exponential increase in the red--shift which signals
the approach to the horizon.  So this quantum limitation seems to probe
the limits of distant experiments to reveal the geometry of a black
hole.

This argument suggests that there is a quantum complementarity between: 
(a) measurements of $v''/v'$ to the precision which detection of Hawking
quanta would imply; and (b) the total energy of the collapsed object. 
Analyzing a thought experiment can never {\it prove } a complementarity, of
course, but in this case the experiment is so natural, and seems so much
to go to the heart of black--hole formation, that we consider the
suggestion a very strong one.  It is also possible to give a
mathematical corroboration of this, using canonical quantization.  This
will be done elsewhere.  

This argument implies that black holes do not radiate.  This is because
detection of Hawking quanta would be equivalent to a measurement of
$v''/v'$, which could not be accomplished without giving rise to a large
spread in the energy of the hole, a macroscopically large spread. 
Hawking radiation is then forbidden by conservation of energy:  there
are no matrix elements available for transition from the state of the
collapsing hole to a state with emitted quanta, as that final state must
involve a too--large spread in the energy of the hole.

This state of affairs is parallel to the familiar one for transitions
in elementary quantum systems.  If a Hamiltonian $H_0$ is perturbed by
a term $\Delta H$ which is considered to give rise to transitions, then
the final $H_0$--width is typically of the order of $\Delta E_{\rm rms}
=\sqrt{}\bigl( \langle (\Delta H)^2\rangle -\langle\Delta H\rangle
^2\bigr)$ (since it is an $H$--eigenstate).  In the present case,
because the collapsed object is very nearly at the black--hole state,
there is no way of mining enough energy from it to give it a width
sufficient to accomodate the production of a Hawking quantum.  All of
the internal energy of the object is red--shifted almost infinitely
relative to infinity, and, since the object is very nearly a black hole,
the potential energy available to infinity is nearly exhausted.

These considerations are essentially new, and are not addressed by
earlier analyses on conservation of energy in the Hawking process.
Those have been of two sorts.  First, as Hawking pointed out, quantum
field theory predicts a negative flux of energy across the event horizon
which counterbalances the energy carried off by Hawking quanta. 
While true, this simply does not address what accomodation the quantum
state of the collapsing matter must make to the emission of the
radiation.  In the analogy of the previous paragraph, it is the
statement that $\langle\Delta H\rangle =0$.

The second sort of energy conservation which has been considered has
been a semi--classical approximation, $R_{ab}-(1/2)Rg_{ab}=-8\pi G
\left( T_{ab}^{\rm classical} +\langle {\widehat T}_{ab}^{\rm
ren}\rangle\right)$.  However, such approximations have as their
hypothesis that the quantum character of corrections can be ignored.

To summarize:  We have used a simple physical model to investigate the
nature of the effects of the quantum back--reaction on the Hawking
process.  This model strongly suggests that there is a quantum
complementarity forbidding simultaneous measurement of (a) that aspect
of the geometry of space--time controlling the production of Hawking
quanta and revealed by their detection, 
and (b) the energy of the collapsing object.  This
complementarity becomes significant at a scale
$$v'\sim (m_{\rm Pl}/m)^2\, ,\eek$$
where $v'(u)$ is the red--shift factor for massless particles arriving
at retarded time $u$ (from equation (\scd )).  
This point is rapidly reached in the collapse
process, and beyond it the emission of Hawking quanta is forbidden by
energy conservation.

\ack
In this Letter, I have doubted that Hawking's prediction of thermal
radiation from black holes is correct.  Yet I hope that I have made my
debt to, and admiration for, his work clear.

I thank Ted Jacobson for bringing the problem of
trans--Planckian modes to my attention, and for patient answers to
questions.

\references

\refjl{Balbinot, R and Fabbri, A 1999}{Nucl. Phys.}{B459}{112}

\refjl{Belinski, V A 1995}{\PL}{A209}{13}

\refjl{Brout, R, Massar, S, Parentani, R, and Spindel, Ph
1995}{\PR}{D52}{4559}

\refjl{Davies, P C W and Fulling, S A 1977}{Proc. R. Soc.
Lond.}{A356}{237}

\refjl{Fredenhagen, K and Haag, R 1990}{\CMP}{127}{273--284}

\refjl{Fulling, S A and Davies, P C W 1976}{Proc. R. Soc.
Lond.}{A348}{393}

\refjl{Hawking, S W 1974}{Nature}{248}{30}

\refjl{Hawking, S W 1975}{\CMP}{43}{199--220}

\refjl{Jacobson, T 1991}{\PR}{D44}{1731}

\refjl{Jacobson, T 1993}{\PR}{D48}{728}

\refjl{Kummer, W and Vassilevich, D V 1999}{Annalen
Phys.}{8}{801--827}

\refbk{Penrose R and Rindler W 1984--6}{Spinors and
Space--Time}{(Cambridge:  University Press)}

\refjl{Rovelli, C 1996}{\PRL}{14}{3288}

\refbk{Schweber S 1961}{An Introduction to Relativistic Quantum Field
Theory}{(Evanston, Illinois and Elmsford, New York:  Row, Peterson and
Company)}

\refjl{Strominger, A and Vafa, G 1996}{\PL}{B379}{99}

\refjl{Unruh, W G 1995}{\PR}{D51}{2827}

\Figures

\figure{Conformal diagram showing the experiment to measure the
fractional acceleration $v''/v'$.  The collapsing body is to the left
(bounded by the dotted line).
Massless particles of given frequency are directed into the body, and
ratios of their red--shifts are measured in the distant future.  As the
event horizon (represented by the dashed line) is approached, the
red--shifts increase exponentially, necessitating exponentially
increasing initial frequencies.}

\epsfbox{figure1}

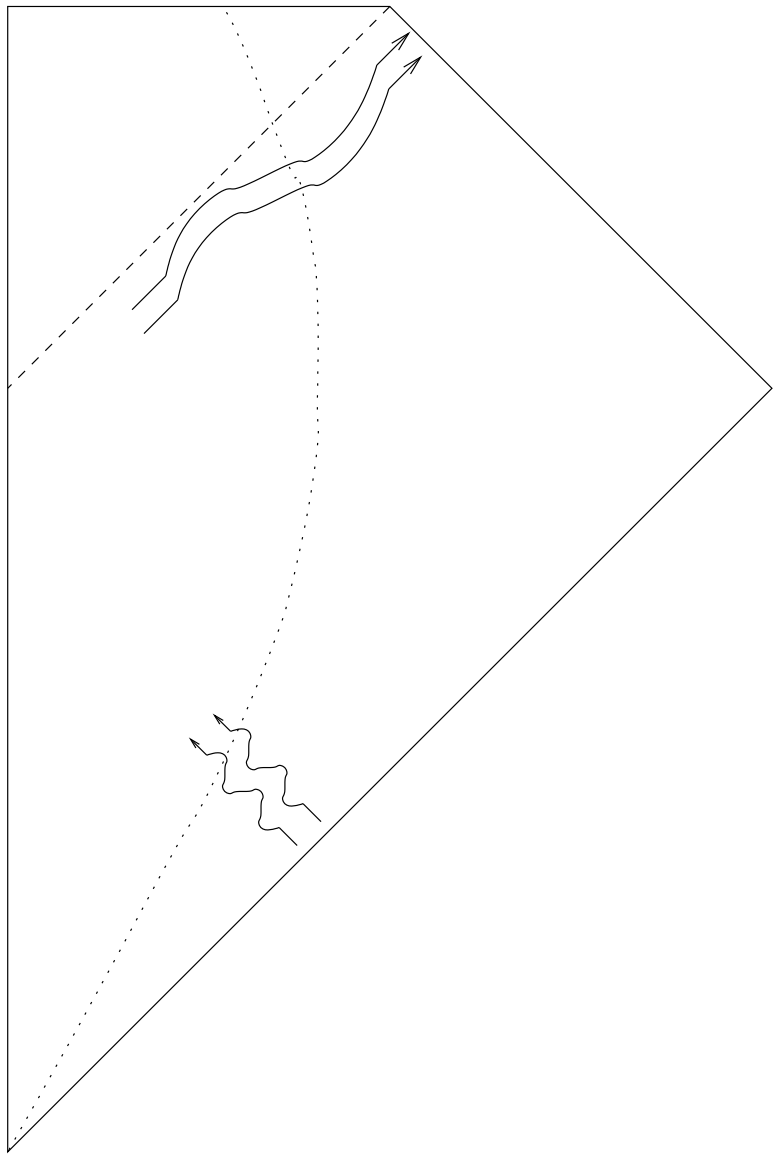
\bye